\documentclass[prl, twocolumn, superscriptaddress, aps]{revtex4-1}

\usepackage[dvips]{graphicx}
\usepackage{color,epsfig}
\usepackage{amsmath,amssymb}
\usepackage{graphicx}

\begin{document}

\title{Selective enhancement of topologically induced interface states in a dielectric resonator chain}

\author{Charles Poli}
\affiliation{Department of Physics, Lancaster University, Lancaster, LA1 4YB, United Kingdom}

\author{Matthieu Bellec}
\affiliation{Laboratoire de Physique de la Mati\`ere Condens\' ee, CNRS UMR 7336, Universit\'e Nice Sophia Antipolis, 06100 Nice, France}

\author{Ulrich Kuhl}
\affiliation{Laboratoire de Physique de la Mati\`ere Condens\' ee, CNRS UMR 7336, Universit\'e Nice Sophia Antipolis, 06100 Nice, France}

\author{Fabrice Mortessagne}
\affiliation{Laboratoire de Physique de la Mati\`ere Condens\' ee, CNRS UMR 7336, Universit\'e Nice Sophia Antipolis, 06100 Nice, France}

\author{Henning  Schomerus}
\affiliation{Department of Physics, Lancaster University, Lancaster, LA1 4YB, United Kingdom}

\maketitle
\textbf{The recent realization of topological phases in insulators and superconductors has advanced the quest for robust quantum technologies. The prospects to implement the underlying topological
features controllably has given incentive to explore optical platforms for analogous
realizations. Here we realize a topologically induced defect state in a chain of dielectric microwave resonators and show that the functionality of the system can  be enhanced by supplementing
topological protection with non-hermitian symmetries that do not have an electronic counterpart. We draw on a characteristic topological feature of the defect
state, namely, that it breaks a sublattice symmetry.
This isolates the state from
losses that respect parity-time symmetry, which
enhances its visibility relative to all other states both in the frequency and in the time domain.
This mode selection mechanism naturally carries over to a wide range of topological and parity-time symmetric optical platforms, including couplers, rectifiers and lasers.}

\begin{figure}[t]
\centering
\includegraphics[width=86mm]{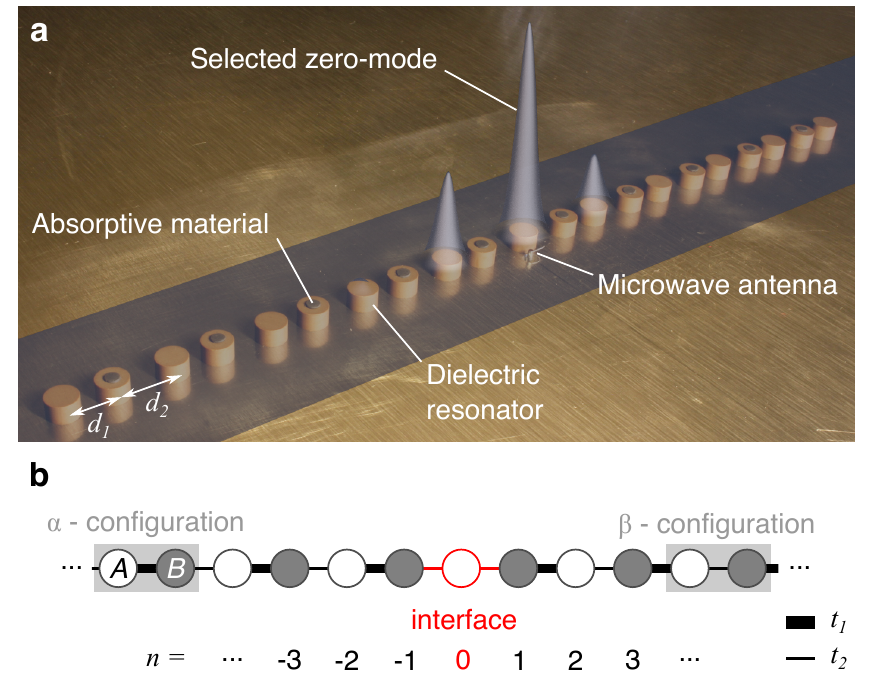}\label{Fig_Exp}
\caption{ \textbf{Realization of a topological defect in a microwave resonator  chain.} \textbf{a.} Picture of the experimental microwave
realization of the complex Su-Schrieffer-Heeger (SSH) chain. The lattice is composed of 21 identical coupled dielectric
cylindrical resonators (5 mm height, 8 mm diameter, and a refractive index of 6) sandwiched between
two metallic plates (note that the top plate is not shown). To implement dimer chains, the resonators
are separated by spacings $d_1$ or $d_2$ with $d_1<d_2$, i.e. couplings $t_1>t_2$. A central
dimerization defect is introduced by repeating the spacing $d_2$.
The defect creates an interface state at zero energy, a zero-mode, whose visibility is enhanced by
means of absorptive patches placed on one of the two sublattices. The resulting wavefunction intensity
is superimposed onto the chain. \textbf{b.} Schematic of the complex SSH chain, with $A$ and
$B$-sublattices indicated in white and gray respectively. The strong (weak) coupling strength is
represented by a thick (thin) line. In our system, the couplings can be controlled by varying the
resonator spacings. The topologically induced zero-mode appears at the interface (red) between
$\alpha$ configuration (with strong intradimer coupling) and $\beta$ configuration (with weak
intradimer coupling).}
\end{figure}

Topological photonics \cite{Lu14} aims to implement robust optical modes by mirroring the interference and interaction effects that drive  condensed matter into topologically protected phases \cite{Has10, Qi11}.
A key element for the intended topological functionality are robust confined states that form at interfaces
between regions with topologically distinct band structures.
For electromagnetic waves, this can be
realized in two dimensions by breaking symmetries in analogy to the quantum Hall effect~\cite{Wan09,
Fan12, Rec13a} or the quantum spin Hall effect~\cite{Haf11, Kha12, Haf13}, while in one dimension one simply
can employ lattice modulations~\cite{Mal09, Kra12, Kit12}.  Concerning the latter setting, a minimal one-dimensional model with a topological band structure is a chain of sites with alternating
couplings (i.e., a dimer chain). This model was originally introduced in an electronic context
by Su, Schrieffer and Heeger (SSH)~\cite{Su79} to describe fractionalized charges in polyacetylene,
which appear in the presence of a dimerization defect. Photonic systems provide a versatile platform
to realize analogies of this situation.
The topological
defect state has been observed in a quantum walk scenario \cite{Kit12}, while a dimer chain with a
non-topological defect has been realised in a waveguide array~\cite{Kei13}.

In absence of the defect, this dimer model has attracted independent attention because it provides a natural platform for gain-loss distributed systems displaying a so-called parity-time  $(\mathcal{PT})$ symmetry~\cite{Guo09, Rut10, Fen12, Reg12, Eich13}.
The topological features of the $\mathcal{PT}$-symmetric variant of the chain without a defect has been discussed in
\cite{Rud09,Ram12}, while two recent experiments have exploited spontaneous $\mathcal{PT}$-symmetry breaking for mode selection in a laser \cite{Feng14, Hod14}. This relies on a mechanism where two modes with real
frequencies coalesce and bifurcate in a strongly and weakly amplified mode \cite{Chong11,Sch10,Lon10}.

The intrinsic robustness of
topologically induced states raises the general question whether they can be controlled and modified independently of the other states in the system. It is then natural to consider whether non-hermitian effects without an electronic analog, such as embodied by the losses and gain in $\mathcal{PT}$-symmetric systems, may be of any help in the photonic setting.

Here, we demonstrate the selective
control and enhancement of the topologically induced state in the SSH  chain in a one-dimensional microwave
set-up. We draw on the passive variant of non-hermitian
$\mathcal{PT}$-symmetry~\cite{Guo09, Fen12, Eich13} and implement the
chain in presence of the defect and localized absorptive
losses by means of a
set of identical coupled dielectric resonators placed in a microwave cavity~\cite{Bel13b, Lau07, Bel13a, Fra13}. The defect state explicitly  breaks the $\mathcal{PT}$ symmetry  \cite{Ryu02,Sch13b}; this topologically induced anomaly further simplifies the mode competition. The state can then be isolated from losses affecting all other
modes in the system, which enhances its visibility in the temporal evolution of a pulse, even in presence of structural disorder.   As the explicit symmetry breaking is a general characteristic feature of topologically induced interface states, our results transfer to a wide range of settings.
Besides its relevance for mode
guiding and filtering, as well as rectifiers and couplers exploiting passive $\mathcal{PT}$ symmetry, this mechanism also lays the conceptual ground for selecting a topologically induced state in mode competition in active variants of the symmetry,
tying it to the topical problem of gain-loss enabled lasing.

\section*{Results}

\textbf{Realization of dimer chains by coupled resonators.} Fig.~\ref{Fig_Exp}a depicts a chain
of 21 microwave resonators with a central dimerization defect.
We establish a one-dimensional tight-binding regime~\cite{Bel13b}, where the electromagnetic field is
mostly confined within the resonators. For an isolated resonator, only a single mode is important in a
broad spectral range around the bare frequency $\nu_\mathrm{b}=6.65$\,GHz. This mode spreads out evanescently,
so that the coupling strength can be controlled by adjusting the separation distance between the
resonators~\cite{Bel13b}. The resulting system can be described by the following tight-binding
equations:
\begin{eqnarray}
(\alpha)\quad&&
\begin{cases}
(\nu -\nu_n)\psi_{n} = t_2 \psi_{n-1}+t_1  \psi_{n+1}, & n =-2,-4,\ldots, \\
(\nu -\nu_n)\psi_{n} = t_1 \psi_{n-1}+t_2  \psi_{n+1}, & n =-1,-3,\ldots,
\end{cases}
\nonumber
\\
\hspace{-1cm}\parbox{2cm}{interface} 
&&\begin{cases}
(\nu -\nu_0)\psi_{0} = t_2 (\psi_{-1}+\psi_{1}),\qquad  & n=0,
\end{cases}
\label{Eq_psi}
 \\
(\beta)\quad&&\begin{cases}
(\nu -\nu_n)\psi_{n} = t_2 \psi_{n-1}+t_1  \psi_{n+1}, & n =1,3,\ldots, \\
(\nu -\nu_n)\psi_{n} = t_1 \psi_{n-1}+t_2  \psi_{n+1}, & n =2,4,\ldots,
\end{cases}
\nonumber
\end{eqnarray}
where $n$ enumerates the resonators, with $n=-10,-9,-8,\ldots 10$ and $n=0$ for the central site (see
Fig.~\ref{Fig_Exp}b). The mode amplitude in the $n$th resonator is given by $\psi_{n}$, $t_1$ and
$t_2$ denote the alternating nearest-neighbor coupling strengths, while $\nu_n$ is the resonance
frequency of the $n$th isolated resonator. Without absorption, the resonance frequencies are equal to
the bare frequency, $\nu_n=\nu_\mathrm{b}$. Absorption is introduced on selected sites by depositing elastomer
patches on top of the dielectric cylinders (see Fig.~\ref{Fig_Exp}a). The losses give rise to a
complex resonance frequency $\nu_n=\nu_\mathrm{b}'-i\gamma$,
and also shift the real part of the bare frequency to $\nu_\mathrm{b}'\approx\nu_\mathrm{b}-\gamma$. In our experiments,
$\gamma \simeq 40$\,MHz,
while the separations $d_1=12$\,mm and $d_2 = 15$\,mm correspond to couplings $t_1 = 37.1$\,MHz and
$t_2 = 14.8$\,MHz, respectively.

\begin{figure}[h]
\centering
\includegraphics[scale=0.9]{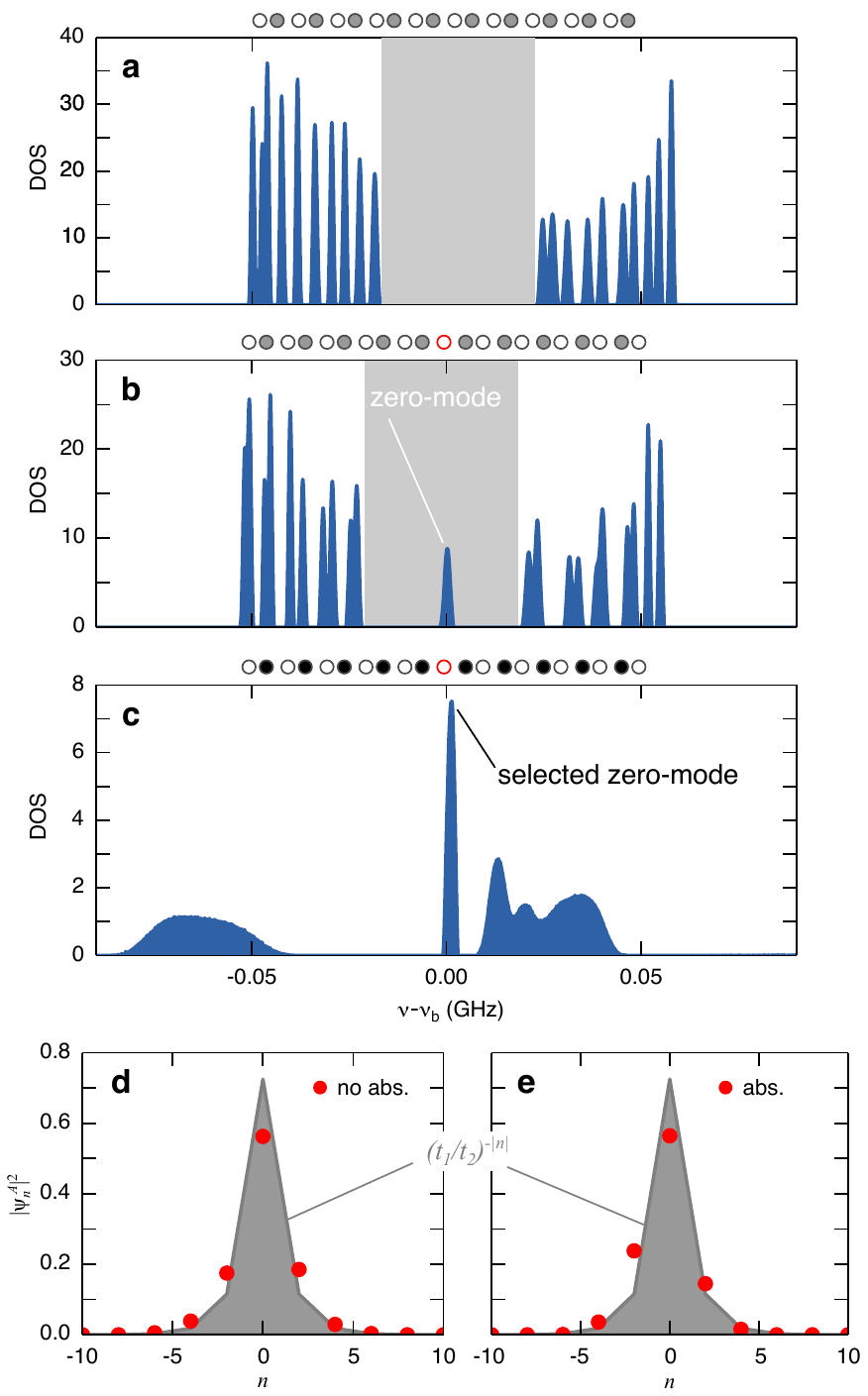}\label{Fig_Spectra} 
\caption{\textbf{Experimental density of states (DOS) and zero-mode profiles of SSH chains with and without absorption.} \textbf{a.} DOS obtained for a defectless SSH chain (no interface, see inset) for separation distances $d_1=12$\,mm ($t_1=37.1\,$MHz) and $d_2=15$\,mm ($t_1=14.8\,$MHz). The reference frequency is the bare frequency of an isolated resonator $\nu_b=6.65\,$GHz. Two bands separated by a gap (gray zone) are observed. \textbf{b.} DOS obtained for a SSH chain with a dimerization defect (inset, red circle). A zero-mode appears in the band gap. \textbf{c.} DOS obtained for a complex SSH chain dimerization defect (inset, red circle) and absorption located on $B$ sites (inset, black filled circles). While all the extended modes experience losses, the zero-mode is preserved. (Note that the ordinate axes scales are different). \textbf{d.} Experimental (red dots) and theoretical (gray region)  intensity profile of the zero-mode on the $A$ sites, for the chain without absorption. The intensity is zero on $B$-sites, and the total intensity is normalized to unity. \textbf{e.} Intensity profile of the zero-mode in presence of losses on $B$-sites.}
\end{figure}

\textbf{Defect-free chains.} As a reference situation, we first consider a defectless SSH chain
without absorption. As depicted in Fig.~\ref{Fig_Spectra}a, the density of states (DOS) is
characterized by a band structure with two bands separated by a finite gap of size
$2|t_1-t_2|=45$\,MHz  (gray zone). The extended states occupy bands in the range
$\nu_\mathrm{b}-t_1-t_2<\nu<\nu_\mathrm{b}-|t_1-t_2|$ and $\nu_\mathrm{b}+|t_1-t_2|<\nu<\nu_\mathrm{b}+t_1+t_2$. The DOS is not affected
when the values of the couplings $t_1$ and $t_2$ are interchanged. Nevertheless, a topological
distinction between these two situations (called hereafter $\alpha$ and $\beta$ configurations) can be
captured by a winding number associated to the Bloch wave functions (see \cite{Ryu02,Sch13b}, Supplementary Note 1 and Supplementary Figs. 1 and 2).
An interface between both configurations takes the form of a dimerization defect where two consecutive
couplings are identical (red zone in Fig.~\ref{Fig_Exp}b and Fig.~\ref{Fig_Spectra}b). The topological
distinctiveness of the two phases leads to the formation of an exponentially localized midgap state at
$\nu=\nu_\mathrm{b}$. The corresponding wavefunction can be read off from equation~(\ref{Eq_psi}), and takes
the form $\psi_n=(-t_1/t_2)^{-|n|/2}$ for even $n$ and $\psi_n=0$ for odd $n$. The midgap state is
therefore confined to the sublattice with even index, which we will call the $A$-sublattice, while the
sites with odd index are called the $B$-sublattice. The complementary state on the $B$-sublattice
increases exponentially and is incompatible with the boundary conditions.

\textbf{Dimerization defect.} Fig.~\ref{Fig_Spectra}b shows the DOS measured for a 21-resonator SSH
chain with a central dimerization defect, still without absorption. Twenty-one modes are observed
within the spectral range of interest. Of these, 20 modes are extended over the whole system. These
modes group into two sets of 10 and correspond to the upper and the lower band of the infinite dimer
chain. The bands remain separated by a 45\,MHz gap. The topologically induced mode clearly sits the
middle of the gap, at frequency $\nu_\mathrm{b}$. We find that the intensity of the midgap state belonging to
the $B$-sublattice is zero within experimental resolution, and thus  confirm that the wavefunction is
confined to the $A$-sublattice. The corresponding wavefunction intensity profile pertaining to the $A$
sites is depicted in Fig.~\ref{Fig_Spectra}d (red dots). As expected, the intensity decays according
to an exponential  profile given by the theoretical result (shown in gray).

\noindent \textbf{Selective enhancement.} We now set out to enhance the visibility of the
topologically induced state. Our approach rests on the realization that the topological features of
the system extend to a staggered configuration of losses, obtained by depositing absorptive material
on all $B$ sites \cite{Sch13b}. Both in the $\alpha$ and in the $\beta$ configuration, the
tight-binding system then still possesses a passive $\mathcal{PT}$-symmetry, given by a reflection
($\mathcal{P}$) at a point in the middle of a dimer, which maps the passive $A$ sites onto the lossy
$B$ sites but leaves the couplings unchanged. This mapping corresponds to a time-reversal operation
($\mathcal{T}$) up to a constant complex frequency shift $i\gamma$. As a consequence of this symmetry,
all extended states are uniformly suppressed by the losses (their complex resonance frequencies all
acquire the same imaginary part $-\gamma/2)$. As it only lives on the $A$-sublattice,
the topologically induced interface state manifestly breaks the $\mathcal{PT}$-symmetry.  In
consequence,  it does not have a complex-conjugate partner, and can be manipulated independently of
all the other states in the system (which is not the case in situations where the
$\mathcal{PT}$-symmetry in the chain is only spontaneously broken~\cite{Rud09,Ram12}). Thus, the
midgap state remains pinned at $\nu=\nu_\mathrm{b}$ and is unaffected by the losses (see Supplementary Note 1).

\begin{figure}[h]
\centering
\includegraphics[width=86mm]{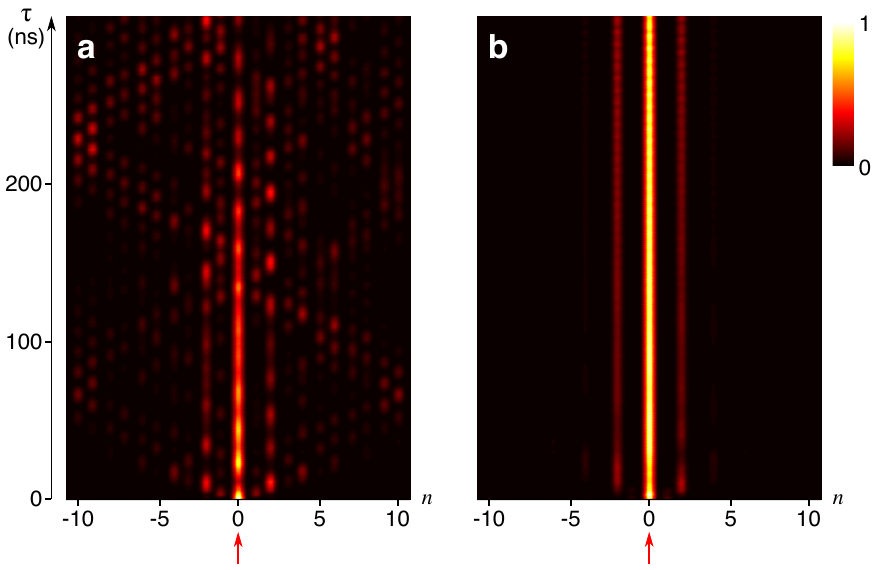}\label{Fig_Propagation}
\caption{ \textbf{Selective enhancement of the zero-mode.} \textbf{a.} Time
evolution of the field intensity (normalized at each time step and color coded by brightness) in the SSH chain without absorption.
The initial excitation is localized at the defect site at the center of the system (indicated by the red
arrow). All modes are participating in the propagation. \textbf{b.} Time evolution of the field
intensity in the complex SSH chain with absorption on $B$-sites. The zero-mode is enhanced and dominates in
the propagation.}
\end{figure}

The spectral analysis presented in Fig.~\ref{Fig_Spectra}c shows that the extended modes shift
downwards in frequency and become broadened, while the overall spectral weight in the resulting
continuous bands is reduced. These features are consistent with the frequency dressing on the $B$
sites,  $\nu_\mathrm{b}\rightarrow \nu_\mathrm{b}-\gamma-i\gamma$. In contrast, the peak in the density of states
associated to the zero-mode remains fixed at the bare frequency $\nu_\mathrm{b}$, and its height and width is
almost unchanged. As shown in Fig.~\ref{Fig_Spectra}e this mode remains well confined to the
$A$-sublattice, and still displays an exponential intensity profile as one moves away from the defect
site. Under the same conditions, non-topological defect states hybridize, thereby degrading
their properties, see Supplementary Note 2 and Supplementary Figs. 3 and 4.

The spectral analysis implies that the interface state is insensitive to the losses distributed on the
$B$ sites. It then should become the predominant mode during the time evolution. To illustrate this
feature, Fig.~\ref{Fig_Propagation} shows the time evolution for both non-absorptive (a) and
absorptive (b) cases, corresponding to a pulse launched on the defect site. Without absorption,
diffraction and interferences spoil the propagation of the interface state, which cannot be discerned
after 250 ns. On the contrary, adding losses drastically enhances the visibility of the topologically
induced mode, which then dominates the propagation without any degradation.

\begin{figure}[h]
\centering
\includegraphics[width=86mm]{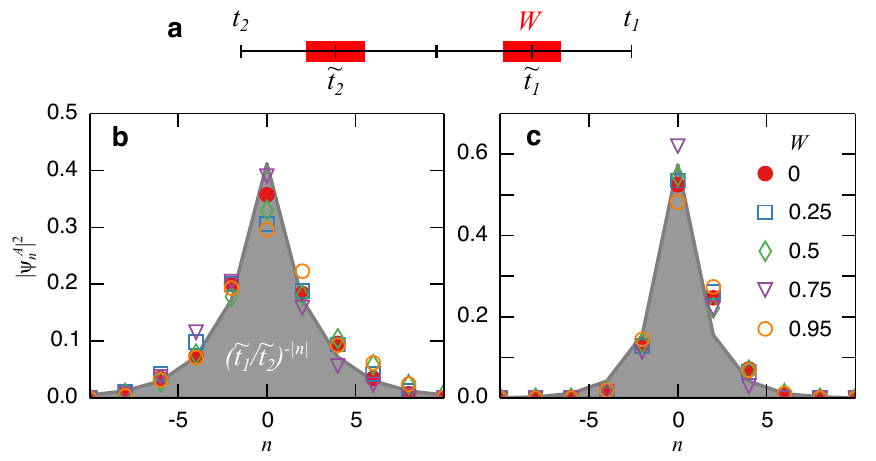}\label{Fig_Profiles}
\caption{ \textbf{Robustness of the zero-mode against structural disorder.} \textbf{a.} Schematic of
the randomized coupling strengths $t_1$, $t_2$ used in the experiments. The disorder strength $W$ (red bar) is chosen to preserve the
topological structure of the chain ($t_1>t_2$). \textbf{b.} Zero-mode intensity profiles (normalized to
1) on $A$ sites for the disordered chain without losses for different values of $W$. The gray shaded
curve follows the exponential profile $(\tilde{t}_1/\tilde{t}_2)^{-|n|}$. \textbf{c.} Intensity
profiles in the presence of losses on the $B$-sites. The  gray  shaded curve is obtained for an
effective coupling ratio $(\tilde{t}_1/\tilde{t}_2)_{\mathrm{eff}} = 1.92$. The topologically induced
interface state is seen to be robust against the structural disorder, both with and without absorption.}
\end{figure}

\noindent \textbf{Robustness against structural disorder.}  To probe the topological protection
property of the midgap state, we introduce structural disorder by randomly distributing the intersite
separations. In the tight-binding description, this corresponds to a random modification of the
coupling strengths. We preserve the dimer structure by defining the couplings as $t_{1,2}^\mathrm{r} =
\tilde{t}_{1,2} + \tilde{t} (W/2) \xi$, where $\tilde{t}_1=(3t_1+t_2)/4=31.5$\,MHz,
$\tilde{t}_2=(t_1+3t_2)/4= 20.4$\,MHz and $\tilde{t} = (t_1-t_2)/2=11.1$\,MHz. Here $W$ is the
disorder strength varying from 0 to 0.95 and $\xi$ is a random number uniformly distributed in the
interval [-1, 1] (see Fig.~\ref{Fig_Profiles}a). As shown in Fig.~\ref{Fig_Profiles}b, for different
values of $W$ the zero-mode intensities exhibit a very similar profile: an approximately exponential
decay on the $A$ sites and insignificant intensity on the $B$ sites.  Even though we do not perform
any averaging over disorder realizations, the experimental profile is in remarkable agreement with the
simple exponential $(\tilde{t}_1/\tilde{t}_2)^{-|n|}$. This robustness in presence of
disorder arises from the sublattice (chiral) symmetry and is further enhanced by the existence of a
finite gap. Note that as $\tilde{t}_1/\tilde{t}_2 = 1.54  < t_1/t_2 = 2.51 $ the present zero-modes
are more extended compared to the situation considered previously (Fig.~\ref{Fig_Spectra}d). When
absorption is added on $B$ sites, we observe in Fig.~\ref{Fig_Profiles}c that the robustness of the
mode persists. Due to the resonance frequency shift of the lossy resonators, the couplings are now
slightly changed to an effective coupling ratio $(\tilde{t}_1/\tilde{t}_2)_{\mathrm{eff}} = 1.92$,
which is taken into account in the theoretical profile (gray shaded curve).

\section*{Discussion}
The loss-induced selective enhancement of the topological interface state observed in this work
exploits the unique structure of the zero-mode wavefunction, which is confined to a sublattice. Such
sublattice-symmetry breaking is common in topologically induced states; for example, it also occurs
for the 0th Landau level in the quantum Hall effect of massless relativistic particles~\cite{Jac84},
as well as in inhomogeneously strained graphene \cite{Gui09} and photonic analogues of deformed
honeycomb  lattices \cite{Rec13b,Sch13a}. The combination of topological constraints and passive
$\mathcal{PT}$-symmetry therefore provides a generic concept which may be exploited in different
settings. This also extends to atom-optical systems, where the defectless version of the passive SSH
model has recently been realized in an optical lattice \cite{Ald13,Ruo02}, while the nonpassive case has
been discussed in theoretical work \cite{Bar13b}. We note that it is also possible to selectively
suppress the interface state, which can be achieved either by placing the losses on the other
sublattice or by interchanging the couplings (topological phases) on both sides of the defect. The
enhancement or suppression of the state can therefore be used to detect the relative size of two
coupling strengths. It is also attractive to replace the losses by amplification, e.g., in a
layered structure of materials with different thickness, which could be used to realize a laser with a
much simplified mode competition, or by nonlinear effects as occurring, e.g., in chains of
coupled quantum dot or quantum well exciton polaritons \cite{Wal13}.

\section*{METHODS}

\noindent \textbf{Microwave realization of tight-binding systems}. The experimental setup is designed to realize a microwave system that is well approximated by a nearest-neighbor tight-binding description~\cite{Bel13b}. The sites of the lattice are occupied by dielectric microwave resonators with a
cylindrical shape (Temex-Ceramics, E2000 series: 5
mm height, 8 mm diameter, and a refractive index of 6). The resonance frequency of an isolated
resonator $\nu_\mathrm{b}$ is around $6.65$\,GHz and corresponds to the on-site energy of atoms in a
tight-binding model. The dielectric cylinders are coupled by the evanescent electromagnetic field, the
corresponding coupling strength $t$ between two resonators is well described by a tight-binding-like
hopping term; $t$ depends on the separation $d$ between resonators.
Via a reflection measurement, one has access, at each site, to the local density of states and to the
wavefunction intensity associated to each eigenfrequency. The density of states (DOS) is obtained by
averaging the local density of states over all resonator positions.  To obtain the time evolution of
the pulse we measure the transmission between a source located at the interface site and a receiver
successively placed at each site position. By performing a Fourier transform, one obtains the temporal
evolution of a pulse initiating from the defect site and propagating into the SSH chain. In all these
experiments, we face an intrinsic on-site disorder of $\sim$ 0.15\%  in the values of $\nu_\mathrm{b}$.

\vspace{1em}
\noindent \textbf{Acknowledgments.} CP and HS acknowledge support of EPSRC via grant EP/J019585/1.

\clearpage

\appendix

\makeatletter
\renewcommand{\theequation}{S.\arabic{equation}}
\renewcommand{\thefigure}{S\@arabic\c@figure}
\makeatother
\setcounter{figure}{0}
\setcounter{equation}{0}

\section*{Supplementary Information}

\noindent \textbf{S1. Theoretical background}
\vspace{0.5em}

\textit{Topological characterization of the dimer chain without losses}.--- The topological features of the band structure of the dimer chain arise from a winding number associated to the Bloch wavefunctions  \cite{Ryu02sup,Del11sup}. Let us first consider the defect-free system in absence of losses. The tight-binding model is then of the form
\begin{eqnarray}
(\nu -\nu_\mathrm{b})\psi_{n}&=&t_{n+1} \psi_{n+1}+t_n  \psi_{n-1}
\label{Eq_psi_general}
\end{eqnarray}
with $t_n=t_a$ when $n$ is odd and $t_n=t_b$ when $n$ is even.  For $0<t_a<t_b$ the system is in the $\alpha$ configuration  while for  $0<t_b<t_a$ it is in the $\beta$ configuration. To distinguish $A$ and $B$ sites we write $\psi_{2n}=\phi_n^{A}$ and $\psi_{2n+1}=\phi_n^{B}$.
Bloch wavefunctions can then be written as $\phi_n^{A,B}=\phi_{A,B}(k)\exp(i k n)$ with $\phi_A=2^{-1/2}$, $\phi_B=\pm 2^{-1/2}(f(k)/|f(k)|)$, where $f(k)=t_a+t_b \exp(i k)$, $k$ being the wavenumber. The corresponding eigenfrequency is $\nu(k)=\nu_\mathrm{b} \pm |f(k)|$, where the plus sign gives the upper band and the minus sign gives the lower band. These bands are shown in the left panel of Supplementary Fig. \ref{Fig_SuppMat_disp}a.

The vector $(\phi^{A}(k),\phi^{B}(k))^T$ can be interpreted as a pseudospin, and thus can be characterized by a polarization vector $\mathbf{P}(k)=(\langle \sigma_x \rangle,\langle\sigma_y\rangle,\langle\sigma_z\rangle)$ composed out of the expectation values of the Pauli matrices. This vector obeys $|\mathbf{P}(k)|=1$. For the Bloch wavefunctions encountered here $\langle\sigma_z\rangle=0$, so that $\mathbf{P}(k)$ is  confined  to the $xy$ plane, with the direction given by  $P_x + i P_y \propto (\nu(k)-\nu_\mathrm{b})f(k)\equiv g(k)$.
As shown in the left panels of Supplementary Fig. \ref{Fig_SuppMat_disp}b,
in the $\alpha$ configuration this vector performs a libration (winding number 0), while in the $\beta$ configuration it performs a rotation (winding number 1).  This difference in the winding number is a topological feature since one cannot smoothly transform between both situations without closing the gap, violating the normalization of $\mathbf{P}(k)$ or leaving the $xy$ plane. In the figure, the blue curve traces out the function $g(k)$ in the complex $P_x+iP_y$ plane. In the $\alpha$ configuration the curve does not encircle the origin, but in the $\beta$ configuration it does.

\begin{figure}[h!]
\includegraphics[width=.8\columnwidth]{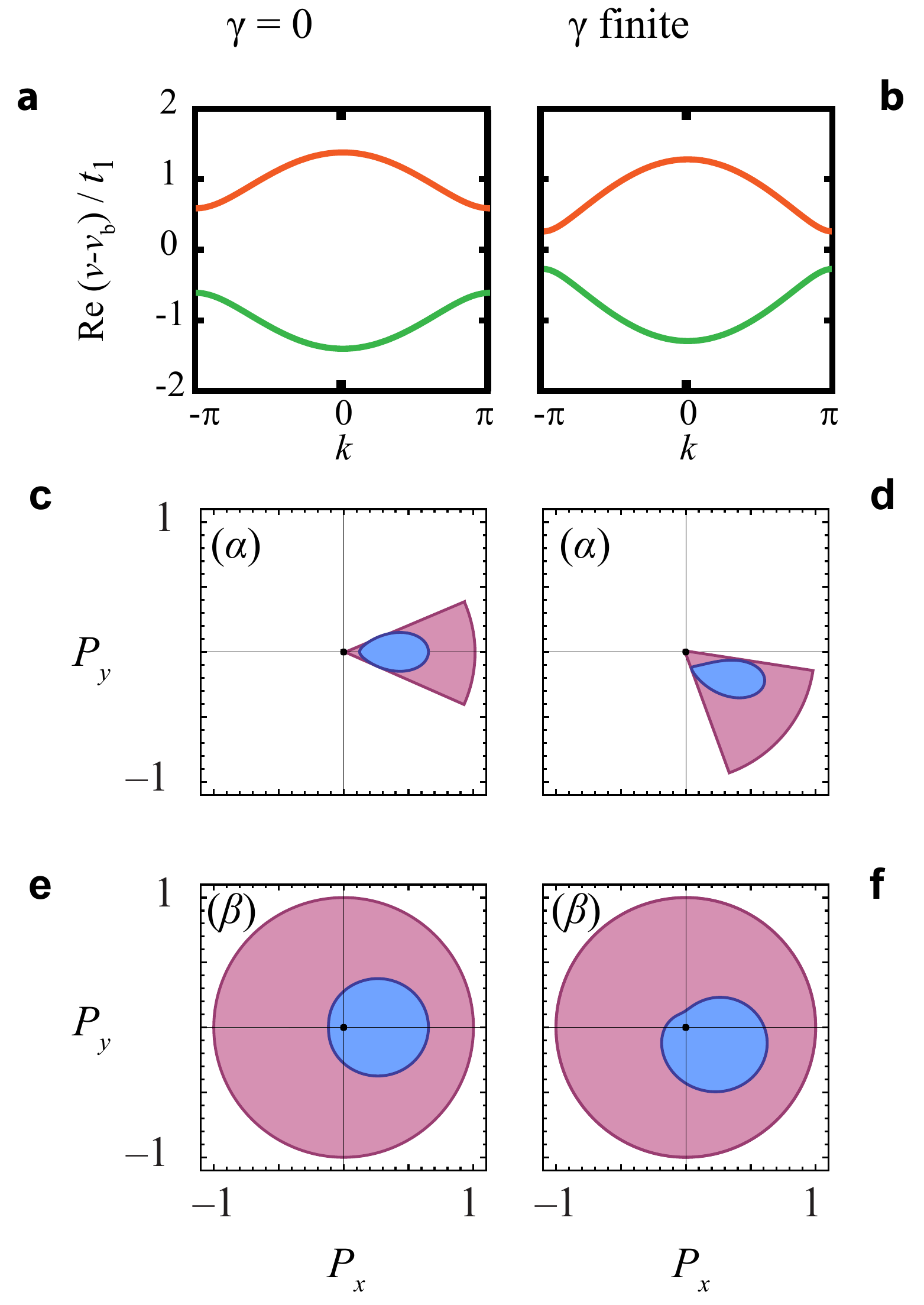}\label{Fig_SuppMat_disp}
\caption{ \textbf{Band structure and topological characterization of the dimer chain without and with losses.}
\textbf{a, b.} Theoretical dispersion relation of the tight-binding chain for couplings $t_a=t_1=37.1$\,MHz, $t_b=t_2=14.8$\,MHz ($\alpha$ configuration), as realized in the experiment. In \textbf{a} the system is loss-free while in \textbf{b} the losses are set to $\gamma \simeq 40$\,MHz, shifting the bare frequency $\nu_\mathrm{b}\to \nu_\mathrm{b}-i\gamma$ on the B sites. The losses reduce the band gap and uniformly shift the extended states into the complex plane, with ${\rm Im}\,\nu(k)=-\gamma/2$. For clarity we neglect the frequency-dressing shift of the real part.  The dispersion in the $\beta$ configuration is identical.
\textbf{c-f.} The red outer curves show the trace of the polarization vector in the $xy$ plane. In the $\alpha$ configuration (\textbf{c,d}), this curve describes a libration while in the $\beta$ configuration (\textbf{e,f}) it describes a rotation. The blue curve shows the function $g(k)=(\nu(k)-\nu_\mathrm{b})f(k)\propto P_x+iP_y$ in appropriately scaled units in the complex $P_x+iP_y$ plane. In the $\alpha$ configuration, the origin lies outside the region encircled by $g(k)$, while in the $\beta$ configuration the origin is enclosed.
}
\end{figure}

When these two topologically distinct configurations are connected as described by Eq.~(1) in the main text, a midgap state appears which is localized at the interface. For a general discussion of topological interface states see  \cite{Has10sup, Qi11sup}.
In the present case, this state can be inferred directly by setting $\nu=\nu_\mathrm{b}$, $\phi_n^{B}=0$, giving the exponential decay $\phi_n^{A}=(-t_1/t_2)^{-|n|}$ on both sides of the interface [i.e., in terms of the amplitudes in Eq.~(\ref{Eq_psi_general}) $\psi_{n}=(-t_1/t_2)^{-|n|/2}$ for even $n$].
The exponential decay corresponds to an imaginary wave number, at which $f(k)=0$. The broken sublattice symmetry results in a polarization vector $\mathbf{P}(k)=(0,0,1)$, pointing out of the $xy$ plane.

\noindent \textit{Topological characterization in presence of losses}.--- Up to a finite threshold $|\gamma|<\gamma_0=2|t_a-t_b|$, these qualitative features survive the introduction of losses on the $B$ sites, corresponding to a shift of the onsite frequency on these sites to $\nu_n=\nu_\mathrm{b}-i\gamma$ \cite{Sch13bsup}. The eigenfrequencies for the Bloch wavefunctions shift to $\nu(k)=  \nu_\mathrm{b}-i \gamma/2 \pm \sqrt{|f(k)|^2-\gamma^2/4}$, meaning that all extended states display the same amount of loss-induced decay. The real part of the dispersion relation  is shown in the right panel of Supplementary Fig. \ref{Fig_SuppMat_disp}a.
The pseudospin vector is given by $\phi_A=2^{-1/2}$, $\phi_B= 2^{-1/2}f(k)/(\nu(k)-\nu_\mathrm{b})$, and the polarization vector is still confined to the $xy$ plane, as shown in  the right panels of Supplementary  Fig. \ref{Fig_SuppMat_disp}b. The characterization of the band structure in terms of winding numbers is thus preserved. In the presence of an interface, the existence of the topologically induced state can again be inferred directly from the tight-binding equations (1) in the main text, giving rise to the same frequency and wavefunction profile as in absence of losses.

For $|\gamma|>\gamma_0$ the interface state persists, but the gap closes, the losses of the extended states become state-dependent, and their spectral weight overlaps with the interface state. Furthermore, the pseudospin vector moves out of the plane, which signifies the breakdown of the topological characterization of the system.\\

\begin{figure}[t]
\includegraphics[width=1.\columnwidth]{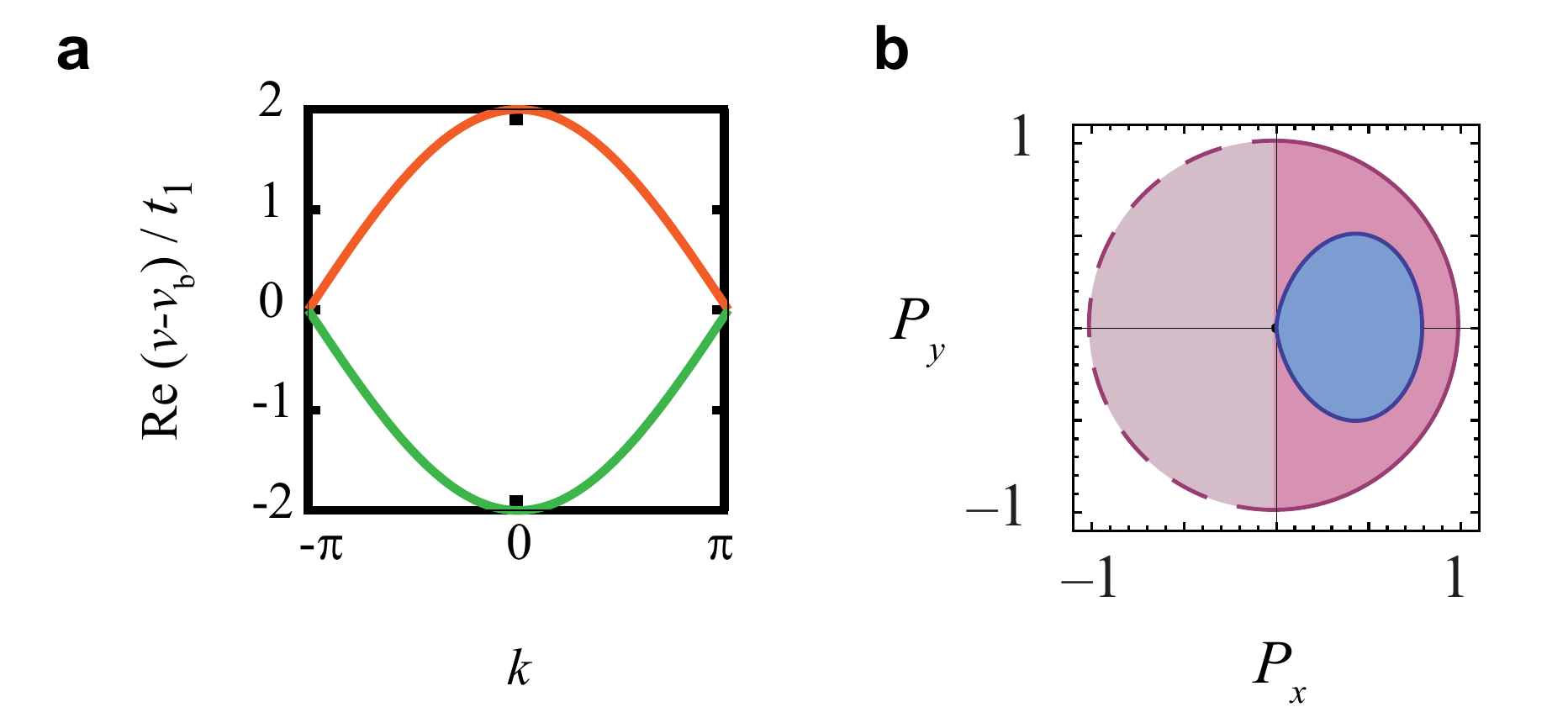}\label{Fig_SuppMat_mono}
\caption{ \textbf{Band structure and topological characterization of the monomer chain.}
\textbf{a.} Theoretical dispersion relation of the tight-binding monomer chain for $t_a=t_b=t_1$, in absence of any losses.
\textbf{b.} As the gap closes, the blue curve showing $g(k)\propto P_x+iP_y$ passes through the origin, and the system is at the switching point between the $\alpha$ and $\beta$ configurations. 
}
\end{figure}

\vspace{1em}
\noindent \textbf{S2. Non-topologically induced defect state}
\vspace{0.5em}

In order to further explore the separate roles of topological protection and spatially distributed losses it is useful to consider a modified set-up in which a defect state is induced via a conventional \textit{non-topological} mechanism. This is achieved by steering the system to the transition point between the $\alpha$ and $\beta$ configurations, which corresponds to a monomer chain with constant site spacing. The gap then closes already in absence of the losses, as the states of the two bands become degenerate at $k=\pm\pi$ (see Supplementary  Fig. \ref{Fig_SuppMat_mono}a).
These bands can be interpreted as two branches of a single band in the unfolded Brillouin zone of the monomer chain.
Out of the degenerate states at $k=\pm\pi$ one can then form superpositions that are confined either to the $A$ or to the $B$-sublattice, with $\mathbf{P}(k)=(0,0,\pm 1)$ just as for the interface state but now achieved for real $k$. Thus, the topological characterization of the system breaks down (see Supplementary  Fig. \ref{Fig_SuppMat_mono}b).

\begin{figure}[b]
\includegraphics[width=0.9\columnwidth]{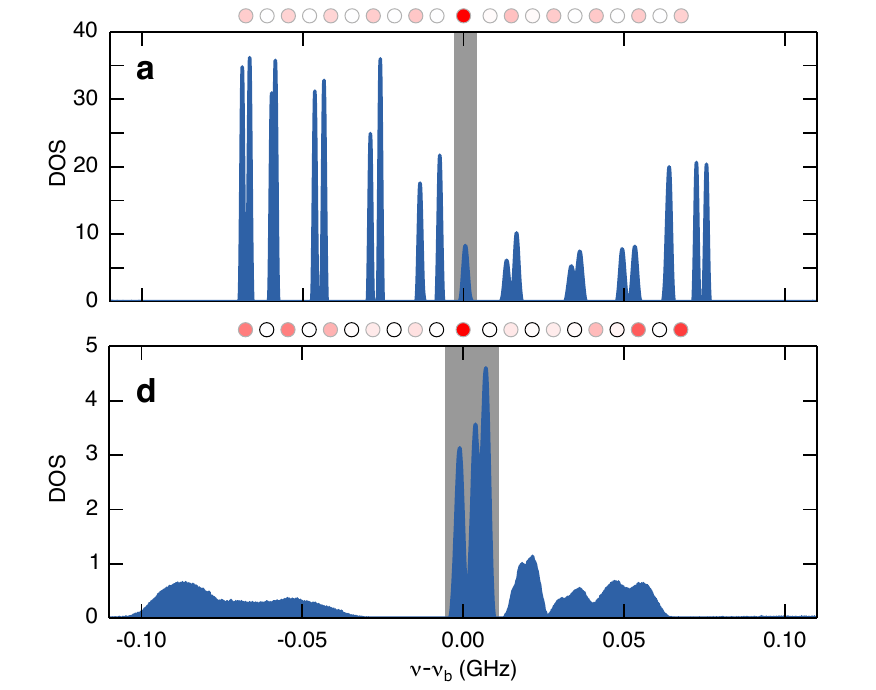}\label{Fig_SuppMat}
\caption{ \textbf{Density of states and zero-mode wavefunction intensities in a monomer chain with a non-topological defect.}  \textbf{a.} Chain of 21 regularly spaced disks ($d_1=12$\,mm) with a central embedded spacing defect ($d_2 = 15$\,mm). This induces a zero mode (corresponding to the gray zone) with an enhanced intensity on the defect site, and a non-decaying tail on the $A$-sublattice (see inset above the panel, which encodes the intensity in red). \textbf{b.} Same as \textbf{a} but with losses on the B sites (indicated in the inset by the black contours). The zero-mode hybridizes, resulting in three broadened peaks of reduced height and an enhanced fluctuating intensity in the tails.}
\end{figure}

\begin{figure}
\includegraphics[width=0.9\columnwidth]{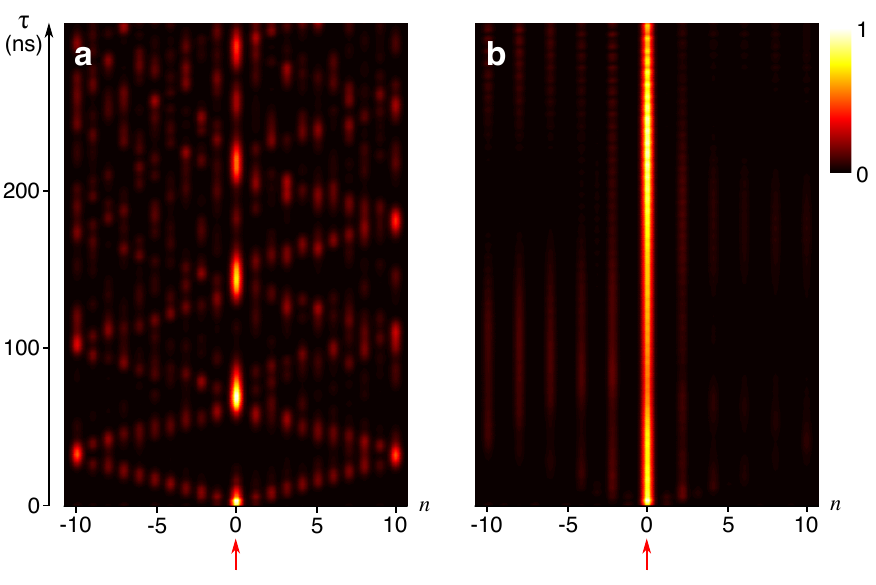}\label{Fig_SuppMat_2} 
\caption{\textbf{Temporal evolution of a pulse in the monomer chains with a non-topological defect.} \textbf{a.} Pulse propagation in absence of losses, with the intensity encoded by brightness (maximal intensity white, vanishing intensity black). \textbf{b.} Same as \textbf{a}, but with losses on $B$ sites. The intensity has been normalized at every time step. The red arrows indicate the site position of the initial excitation.}
\end{figure}

In our experiments, we set the uniform spacing to $d_1 = 12$\,mm (coupling strength $t_1=37.1$\,MHz) and maintain the embedded spacing defect at the position of the central resonator ($d_2=15$\,mm, i.e. $t_2=14.8$\,MHz). The system then still displays a defect mode at the bare frequency $\nu_\mathrm{b}$, but instead within a band gap this mode now lies in the middle of the unfolded band.
This is verified by the experimental spectral analysis depicted in Supplementary Fig. \ref{Fig_SuppMat}a, where the position of the state in the frequency spectrum is highlighted in gray. The corresponding wavefunction intensity (inset) shows that the defect mode is enhanced at the defect, but now possesses a non-decaying tail on the $A$-sublattice.  This concurs with the prediction from the tight-binding model, according to which the tail is given by $|\psi_n|^2=(t_2/t_1)^2 |\psi_0|^2$ ($n \neq 0$  even).

It is interesting to note that in absence of any disorder we still find zero intensity on the $B$-sublattice, as for the topologically induced defect state. The persistence of the sublattice-polarization follows from the chiral symmetry of tight-binding model with a bipartite lattice structure \cite{Lie89sup, Bro02sup}. In the one-dimensional tight-binding chain this symmetry is realized trivially because $A$ sites only couple to $B$ sites and vice versa.
However, as this feature depends on a symmetry it is far less robust than in the case of the topologically induced mode. For example, we find that a small amount of disorder now suffices to induce hybridization with other extended states in the system. In particular, hybridization already occurs when losses are introduced onto the $B$ sites. As shown in Supplementary Fig. \ref{Fig_SuppMat}b, in the density of states several broadened peaks of reduced height then overlap in the region around the bare frequency. The associated spatial intensity distribution (inset) is still large at the defect site, but the weight in the tails increases, especially close to the edges of the system. In the time domain (see Supplementary Figs.~\ref{Fig_SuppMat_2}a and b), the pulse propagation profile again stabilizes when absorption is added, but the intensity on the defect site fluctuates and an oscillating intensity remains clearly visible in the tails. Thus, in absence of topological protection the defect state is not isolated against the losses, making the propagation less robust than for the topologically induced state studied in the paper.


\begin{thebibliography}{10}
\expandafter\ifx\csname url\endcsname\relax
  \def\url#1{\texttt{#1}}\fi
\expandafter\ifx\csname urlprefix\endcsname\relax\def\urlprefix{URL }\fi
\providecommand{\bibinfo}[2]{#2}
\providecommand{\eprint}[2][]{\url{#2}}

\bibitem{Lu14}
\bibinfo{author}{Lu, L.},  \bibinfo{author}{Joannopoulos, J.~D.} \& \bibinfo{author}{Solja\v{c}i\'c, M.}
\newblock \bibinfo{title}{Topological photonics}.
\newblock \emph{\bibinfo{journal}{Nat. Photon.}}
  \textbf{\bibinfo{volume}{8}}, \bibinfo{pages}{821--829}
  (\bibinfo{year}{2014}).

\bibitem{Has10}
\bibinfo{author}{Hasan, M.~Z.} \& \bibinfo{author}{Kane, C.~L.}
\newblock \bibinfo{title}{Topological insulators}.
\newblock \emph{\bibinfo{journal}{Rev. Mod. Phys.}}
  \textbf{\bibinfo{volume}{82}}, \bibinfo{pages}{3045--3067}
  (\bibinfo{year}{2010}).

\bibitem{Qi11}
\bibinfo{author}{Qi, X.-L.} \& \bibinfo{author}{Zhang, S.-C.}
\newblock \bibinfo{title}{Topological insulators and superconductors}.
\newblock \emph{\bibinfo{journal}{Rev. Mod. Phys.}}
  \textbf{\bibinfo{volume}{83}}, \bibinfo{pages}{1057--1110} (\bibinfo{year}{2011}).





\bibitem{Wan09}
\bibinfo{author}{Wang, Z.}, \bibinfo{author}{Chong, Y.} \bibinfo{author}{Joannopoulos, J. D.} \&
\bibinfo{author}{Solja\v{c}i\'c, M.}
\newblock \bibinfo{title}{Observation of unidirectional backscattering-immune
  topological electromagnetic states}.
\newblock \emph{\bibinfo{journal}{Nature}} \textbf{\bibinfo{volume}{461}},
  \bibinfo{pages}{772--775} (\bibinfo{year}{2009}).

\bibitem{Fan12}
\bibinfo{author}{Fang, K.}, \bibinfo{author}{Yu, Z.} \& \bibinfo{author}{Fan,
  S.}
\newblock \bibinfo{title}{Realizing effective magnetic field for photons by
  controlling the phase of dynamic modulation}.
\newblock \emph{\bibinfo{journal}{Nat. Photon.}}
  \textbf{\bibinfo{volume}{6}}, \bibinfo{pages}{782--787}
  (\bibinfo{year}{2012}).

\bibitem{Rec13a}
\bibinfo{author}{Rechtsman, M.~C.} \emph{et~al.}
\newblock \bibinfo{title}{Photonic {F}loquet topological insulators}.
\newblock \emph{\bibinfo{journal}{Nature}} \textbf{\bibinfo{volume}{496}},
  \bibinfo{pages}{196--200} (\bibinfo{year}{2013}).

\bibitem{Haf11}
\bibinfo{author}{Hafezi, M.}, \bibinfo{author}{Demler, E.~A.},
  \bibinfo{author}{Lukin, M.~D.} \& \bibinfo{author}{Taylor, J.~M.}
\newblock \bibinfo{title}{Robust optical delay lines with topological
  protection}.
\newblock \emph{\bibinfo{journal}{Nat. Phys.}} \textbf{\bibinfo{volume}{7}},
  \bibinfo{pages}{907--912} (\bibinfo{year}{2011}).

\bibitem{Kha12}
\bibinfo{author}{Khanikaev, A.~B.}, 
\bibinfo{author}{Mousavi, S.~H.},
\bibinfo{author}{Tse, W.-K.},
\bibinfo{author}{Kargarin, M.},
\bibinfo{author}{MacDonald, A.~H.} \&
\bibinfo{author}{Shvets, G.}
\newblock \bibinfo{title}{Photonic topological insulators}.
\newblock \emph{\bibinfo{journal}{Nat. Mater.}}
  \textbf{\bibinfo{volume}{12}}, \bibinfo{pages}{233--239}
  (\bibinfo{year}{2013}).

\bibitem{Haf13}
\bibinfo{author}{Hafezi, M.}, \bibinfo{author}{Mittal, S.},
  \bibinfo{author}{Fan, J.}, \bibinfo{author}{Migdall, A.} \&
  \bibinfo{author}{Taylor, J.~M.}
\newblock \bibinfo{title}{Imaging topological edge states in silicon
  photonics}.
\newblock \emph{\bibinfo{journal}{Nat. Photon.}}
  \textbf{\bibinfo{volume}{7}}, \bibinfo{pages}{1001--1005}
  (\bibinfo{year}{2013}).

\bibitem{Mal09}
\bibinfo{author}{Malkova, N.}, \bibinfo{author}{Hromada, I.},
  \bibinfo{author}{Wang, X.}, \bibinfo{author}{Bryant, G.} \&
  \bibinfo{author}{Chen, Z.}
\newblock \bibinfo{title}{Observation of optical {S}hockley-like surface states
  in photonic superlattices}.
\newblock \emph{\bibinfo{journal}{Opt. Lett.}} \textbf{\bibinfo{volume}{34}},
  \bibinfo{pages}{1633--1635} (\bibinfo{year}{2009}).

\bibitem{Kra12}
\bibinfo{author}{Kraus, Y.~E.}, \bibinfo{author}{Lahini, Y.},
  \bibinfo{author}{Ringel, Z.}, \bibinfo{author}{Verbin, M.} \&
  \bibinfo{author}{Zilberberg, O.}
\newblock \bibinfo{title}{Topological states and adiabatic pumping in
  quasicrystals}.
\newblock \emph{\bibinfo{journal}{Phys. Rev. Lett.}}
  \textbf{\bibinfo{volume}{109}}, \bibinfo{pages}{106402}
  (\bibinfo{year}{2012}).

\bibitem{Kit12}
\bibinfo{author}{Kitagawa, T.} \emph{et~al.}
\newblock \bibinfo{title}{Observation of topologically protected bound states
  in photonic quantum walks}.
\newblock \emph{\bibinfo{journal}{Nat. Commun.}} \textbf{\bibinfo{volume}{3}}, \bibinfo{pages}{882}  (\bibinfo{year}{2012}).

\bibitem{Su79}
\bibinfo{author}{Su, W.~P.}, \bibinfo{author}{Schrieffer, J.~R.} \&
  \bibinfo{author}{Heeger, A.~J.}
\newblock \bibinfo{title}{Solitons in polyacetylene}.
\newblock \emph{\bibinfo{journal}{Phys. Rev. Lett.}}
  \textbf{\bibinfo{volume}{42}}, \bibinfo{pages}{1698--1701} (\bibinfo{year}{1979}).


\bibitem{Kei13}
\bibinfo{author}{Keil, R.} \emph{et~al.}
\newblock \bibinfo{title}{The random mass dirac model and long-range
  correlations on an integrated optical platform}.
\newblock \emph{\bibinfo{journal}{Nat. Commun.}}
  \textbf{\bibinfo{volume}{4}}, \bibinfo{pages}{1368} (\bibinfo{year}{2013}).


\bibitem{Guo09}
\bibinfo{author}{Guo, A.} \emph{et~al.}
\newblock \bibinfo{title}{Observation of $\mathcal{P}\mathcal{T}$-symmetry
  breaking in complex optical potentials}.
\newblock \emph{\bibinfo{journal}{Phys. Rev. Lett.}}
  \textbf{\bibinfo{volume}{103}}, \bibinfo{pages}{093902}
  (\bibinfo{year}{2009}).

\bibitem{Rut10}
\bibinfo{author}{R{\"u}ter, C.~E.},
\bibinfo{author}{Makris, K.~G.},
\bibinfo{author}{El-Ganainy, R.},
\bibinfo{author}{Christodoulides, D.~N.},
\bibinfo{author}{Segev, M.} \&
\bibinfo{author}{Kip, D.}
\newblock \bibinfo{title}{Observation of parity--time symmetry in optics}.
\newblock \emph{\bibinfo{journal}{Nat. Phys.}} \textbf{\bibinfo{volume}{6}},
  \bibinfo{pages}{192--195} (\bibinfo{year}{2010}).

\bibitem{Fen12}
\bibinfo{author}{Feng, L.} \emph{et~al.}
\newblock \bibinfo{title}{Experimental demonstration of a unidirectional
  reflectionless parity--time metamaterial at optical frequencies}.
\newblock \emph{\bibinfo{journal}{Nat. Mater.}}
  \textbf{\bibinfo{volume}{12}}, \bibinfo{pages}{108--113}
  (\bibinfo{year}{2013}).

\bibitem{Reg12}
\bibinfo{author}{Regensburger, A.},
\bibinfo{author}{Bersch, C.},
\bibinfo{author}{Miri, M.-A.},
\bibinfo{author}{Onishchukov, G.},
\bibinfo{author}{Christodoulides, D.~N.} \&
\bibinfo{author}{Peschel, U.}
\newblock \bibinfo{title}{Parity-time synthetic photonic lattices}.
\newblock \emph{\bibinfo{journal}{Nature}} \textbf{\bibinfo{volume}{488}},
  \bibinfo{pages}{167--171} (\bibinfo{year}{2012}).


\bibitem{Eich13}
\bibinfo{author}{Eichelkraut, T.} \emph{et~al.}
\newblock \bibinfo{title}{Mobility transition from ballistic to diffusive transport in non-Hermitian lattices}.
\newblock \emph{\bibinfo{journal}{Nat. Commun.}} \textbf{\bibinfo{volume}{4}},
  \bibinfo{pages}{2533} (\bibinfo{year}{2013}).


\bibitem{Rud09}
\bibinfo{author}{Rudner, M. S.} \& \bibinfo{author}{Levitov, L. S.}
\newblock \bibinfo{title}{Topological Transition in a Non-Hermitian Quantum Walk}.
\newblock \emph{\bibinfo{journal}{Phys. Rev. Lett.}}
  \textbf{\bibinfo{volume}{102}}, \bibinfo{pages}{065703}
  (\bibinfo{year}{2009}).

\bibitem{Ram12}
\bibinfo{author}{Ramezani, H.}, \bibinfo{author}{Christodoulides, D.~N.},
  \bibinfo{author}{Kovanis, V.}, \bibinfo{author}{Vitebskiy, I.} \&
  \bibinfo{author}{Kottos, T.}
\newblock \bibinfo{title}{$\mathcal{PT}$-symmetric {T}albot effect}.
\newblock \emph{\bibinfo{journal}{Phys. Rev. Lett.}}
  \textbf{\bibinfo{volume}{109}}, \bibinfo{pages}{033902}
  (\bibinfo{year}{2012}).

\bibitem{Feng14}
\bibinfo{author}{Feng, L.},
\bibinfo{author}{Wong, Z.~J.},
\bibinfo{author}{Ma, R.-M.},
\bibinfo{author}{Wang, Y.} \& \bibinfo{author}{Zhang, X.}
\newblock \bibinfo{title}{Single-mode laser by parity-time symmetry breaking}.
\newblock \emph{\bibinfo{journal}{Science}} \textbf{\bibinfo{volume}{346}},
\bibinfo{pages}{972--975} (\bibinfo{year}{2014}).


\bibitem{Hod14} 
\bibinfo{author}{Hodaei, H.}, \bibinfo{author}{Miri, M.-A.}, \bibinfo{author}{Heinrich, M.}, \bibinfo{author}{Christodoulides, D.~N.} \& \bibinfo{author}{Khajavikhan, M.}
\newblock \bibinfo{title}{Parity-time-symmetric microring lasers}.
\newblock \emph{\bibinfo{journal}{Science}} \textbf{\bibinfo{volume}{346}},
\bibinfo{pages}{975--978} (\bibinfo{year}{2014}).


\bibitem{Chong11}
\bibinfo{author}{Chong, Y. D.}, \bibinfo{author}{Ge, L.} \&
  \bibinfo{author}{Stone, A. D.}
\newblock \bibinfo{title}{$\mathcal{PT}$-Symmetry Breaking and Laser-Absorber Modes in Optical
Scattering Systems}.
\newblock \emph{\bibinfo{journal}{Phys. Rev. Lett.}}
  \textbf{\bibinfo{volume}{106}}, \bibinfo{pages}{093902}
  (\bibinfo{year}{2011}).

\bibitem{Sch10}
\bibinfo{author}{Schomerus, H.}
\newblock \bibinfo{title}{Quantum Noise and Self-Sustained Radiation of $\mathcal{PT}$-Symmetric Systems}.
\newblock \emph{\bibinfo{journal}{Phys. Rev. Lett.}}
  \textbf{\bibinfo{volume}{104}}, \bibinfo{pages}{233601}
  (\bibinfo{year}{2010}).

\bibitem{Lon10}
\bibinfo{author}{Longhi, S.}
\newblock \bibinfo{title}{$\mathcal{PT}$-symmetric laser absorber}.
\newblock \emph{\bibinfo{journal}{Phys. Rev. A}}
  \textbf{\bibinfo{volume}{82}}, \bibinfo{pages}{031801(R)}
  (\bibinfo{year}{2010}).

\bibitem{Bel13b}
\bibinfo{author}{Bellec, M.}, \bibinfo{author}{Kuhl, U.},
  \bibinfo{author}{Montambaux, G.} \& \bibinfo{author}{Mortessagne, F.}
\newblock \bibinfo{title}{Tight-binding couplings in microwave artificial
  graphene}.
\newblock \emph{\bibinfo{journal}{Phys. Rev. B}} \textbf{\bibinfo{volume}{88}},
  \bibinfo{pages}{115437} (\bibinfo{year}{2013}).

\bibitem{Lau07}
\bibinfo{author}{Laurent, D.}, \bibinfo{author}{Legrand, O.},
  \bibinfo{author}{Sebbah, P.}, \bibinfo{author}{Vanneste, C.} \&
  \bibinfo{author}{Mortessagne, F.}
\newblock \bibinfo{title}{Localized modes in a finite-size open disordered
  microwave cavity}.
\newblock \emph{\bibinfo{journal}{Phys. Rev. Lett.}}
  \textbf{\bibinfo{volume}{99}}, \bibinfo{pages}{253902}
  (\bibinfo{year}{2007}).

\bibitem{Bel13a}
\bibinfo{author}{Bellec, M.}, \bibinfo{author}{Kuhl, U.},
  \bibinfo{author}{Montambaux, G.} \& \bibinfo{author}{Mortessagne, F.}
\newblock \bibinfo{title}{Topological transition of {D}irac points in a
  microwave experiment}.
\newblock \emph{\bibinfo{journal}{Phys. Rev. Lett.}}
  \textbf{\bibinfo{volume}{110}}, \bibinfo{pages}{033902}
  (\bibinfo{year}{2013}).

\bibitem{Fra13}
\bibinfo{author}{Franco-Villafa\~ne, J.~A.},
\bibinfo{author}{Sadurn\'\i, E.},
\bibinfo{author}{Barkhofen, S.},
\bibinfo{author}{Kuhl, U.},
\bibinfo{author}{Mortessagne, F.} \&
\bibinfo{author}{Seligman, T.~H.}
\newblock \bibinfo{title}{First experimental realization of the {D}irac
  oscillator}.
\newblock \emph{\bibinfo{journal}{Phys. Rev. Lett.}}
  \textbf{\bibinfo{volume}{111}}, \bibinfo{pages}{170405}
  (\bibinfo{year}{2013}).


\bibitem{Ryu02}
\bibinfo{author}{Ryu, S.} \& \bibinfo{author}{Hatsugai, Y.}
\newblock \bibinfo{title}{Topological origin of zero-energy edge states in
  particle-hole symmetric systems}.
\newblock \emph{\bibinfo{journal}{Phys. Rev. Lett.}}
  \textbf{\bibinfo{volume}{89}}, \bibinfo{pages}{077002}
  (\bibinfo{year}{2002}).

\bibitem{Sch13b}
\bibinfo{author}{Schomerus, H.}
\newblock \bibinfo{title}{Topologically protected midgap states in complex
  photonic lattices}.
\newblock \emph{\bibinfo{journal}{Opt. Lett.}} \textbf{\bibinfo{volume}{38}},
  \bibinfo{pages}{1912--1914} (\bibinfo{year}{2013}).


\bibitem{Jac84}
\bibinfo{author}{Jackiw, R.}
\newblock \bibinfo{title}{Fractional charge and zero modes for planar systems
  in a magnetic field}.
\newblock \emph{\bibinfo{journal}{Phys. Rev. D}} \textbf{\bibinfo{volume}{29}},
  \bibinfo{pages}{2375--2377} (\bibinfo{year}{1984}).

\bibitem{Gui09}
\bibinfo{author}{Guinea, F.}, \bibinfo{author}{Katsnelson, M.~I.} \&
  \bibinfo{author}{Geim, A.~K.}
\newblock \bibinfo{title}{Energy gaps and a zero-field quantum {H}all effect in
  graphene by strain engineering}.
\newblock \emph{\bibinfo{journal}{Nat. Phys.}} \textbf{\bibinfo{volume}{6}},
  \bibinfo{pages}{30--33} (\bibinfo{year}{2009}).

\bibitem{Rec13b}
\bibinfo{author}{Rechtsman, M.~C.},	
\bibinfo{author}{Zeuner, J.~M.},	
\bibinfo{author}{T{\"u}nnermann, A.},	
\bibinfo{author}{Nolte, S.},	
\bibinfo{author}{Segev, M.} \&	
\bibinfo{author}{Alexander Szameit}
\newblock \bibinfo{title}{Strain-induced pseudomagnetic field and photonic
  {L}andau levels in dielectric structures}.
\newblock \emph{\bibinfo{journal}{Nat. Photon.}}
  \textbf{\bibinfo{volume}{7}}, \bibinfo{pages}{153--158}
  (\bibinfo{year}{2013}).

\bibitem{Sch13a}
\bibinfo{author}{Schomerus, H.} \& \bibinfo{author}{Halpern, N.~Y.}
\newblock \bibinfo{title}{Parity anomaly and {L}andau-level lasing in strained
  photonic honeycomb lattices}.
\newblock \emph{\bibinfo{journal}{Phys. Rev. Lett.}}
  \textbf{\bibinfo{volume}{110}}, \bibinfo{pages}{013903}
  (\bibinfo{year}{2013}).

\bibitem{Ald13}
\bibinfo{author}{Atala, M.} \emph{et~al.}
\newblock \bibinfo{title}{Direct measurement of the {Z}ak phase in topological
  {B}loch bands}.
\newblock \emph{\bibinfo{journal}{Nat. Phys.}} \textbf{\bibinfo{volume}{9}},
  \bibinfo{pages}{795--800} (\bibinfo{year}{2013}).

\bibitem{Ruo02}
\bibinfo{author}{Ruostekoski, J.},
\bibinfo{author}{Dunne, G.~V.} \&
\bibinfo{author}{Javanainen, J.}
\newblock \bibinfo{title}{Particle number fractionilization of an atomic Fermi-Dirac gas in an optical lattice}.
\newblock \emph{\bibinfo{journal}{Phys. Rev. Lett.}} \textbf{\bibinfo{volume}{88}},
  \bibinfo{pages}{180401} (\bibinfo{year}{2002}).

\bibitem{Bar13b}
\bibinfo{author}{Barnett, R.}
\newblock \bibinfo{title}{Edge-state instabilities of bosons in a topological
  band}.
\newblock \emph{\bibinfo{journal}{Phys. Rev. A}} \textbf{\bibinfo{volume}{88}},
  \bibinfo{pages}{063631} (\bibinfo{year}{2013}).

\bibitem{Wal13}
\bibinfo{author}{Walker, P.~M.} \emph{et~al.}
\newblock \bibinfo{title}{Exciton polaritons in semiconductor waveguides}.
\newblock \emph{\bibinfo{journal}{Appl. Phys. Lett.}}
  \textbf{\bibinfo{volume}{102}}, \bibinfo{pages}{012109}
  (\bibinfo{year}{2013}).

\end{thebibliography}

\begin{thebibliography}{1}
\expandafter\ifx\csname url\endcsname\relax
  \def\url#1{\texttt{#1}}\fi
\expandafter\ifx\csname urlprefix\endcsname\relax\def\urlprefix{URL }\fi
\providecommand{\bibinfo}[2]{#2}
\providecommand{\eprint}[2][]{\url{#2}}

\bibitem{Ryu02sup}
\bibinfo{author}{Ryu, S.} \& \bibinfo{author}{Hatsugai, Y.}
\newblock \bibinfo{title}{Topological origin of zero-energy edge states in
  particle-hole symmetric systems}.
\newblock \emph{\bibinfo{journal}{Phys. Rev. Lett.}}
  \textbf{\bibinfo{volume}{89}}, \bibinfo{pages}{077002}
  (\bibinfo{year}{2002}).

\bibitem{Del11sup}
\bibinfo{author}{Delplace, P.}, \bibinfo{author}{Ullmo, D.} \&
  \bibinfo{author}{Montambaux, G.}
\newblock \bibinfo{title}{{Z}ak phase and the existence of edge states in
  graphene}.
\newblock \emph{\bibinfo{journal}{Phys. Rev. B}} \textbf{\bibinfo{volume}{84}},
  \bibinfo{pages}{195452} (\bibinfo{year}{2011}).

\bibitem{Has10sup}
\bibinfo{author}{Hasan, M.~Z.} \& \bibinfo{author}{Kane, C.~L.}
\newblock \bibinfo{title}{Topological insulators}.
\newblock \emph{\bibinfo{journal}{Rev. Mod. Phys.}}
  \textbf{\bibinfo{volume}{82}}, \bibinfo{pages}{3045--3067}
  (\bibinfo{year}{2010}).

\bibitem{Qi11sup}
\bibinfo{author}{Qi, X.-L.} \& \bibinfo{author}{Zhang, S.-C.}
\newblock \bibinfo{title}{Topological insulators and superconductors}.
\newblock \emph{\bibinfo{journal}{Rev. Mod. Phys.}}
  \textbf{\bibinfo{volume}{83}}, \bibinfo{pages}{1057--1110} (\bibinfo{year}{2011}).

\bibitem{Sch13bsup}
\bibinfo{author}{Schomerus, H.}
\newblock \bibinfo{title}{Topologically protected midgap states in complex
  photonic lattices}.
\newblock \emph{\bibinfo{journal}{Opt. Lett.}} \textbf{\bibinfo{volume}{38}},
  \bibinfo{pages}{1912--1914} (\bibinfo{year}{2013}).

\bibitem{Lie89sup}
\bibinfo{author}{Lieb, E.~H.}
\newblock \bibinfo{title}{Two theorems on the {H}ubbard model}.
\newblock \emph{\bibinfo{journal}{Phys. Rev. Lett.}}
  \textbf{\bibinfo{volume}{62}}, \bibinfo{pages}{1201--1204} (\bibinfo{year}{1989}).

\bibitem{Bro02sup}
\bibinfo{author}{Brouwer, P.~W.}, \bibinfo{author}{Racine, E.},
\bibinfo{author}{Furusaki, A.}, \bibinfo{author}{Hatsugai, Y.}, \bibinfo{author}{Morita, Y.} \& \bibinfo{author}{Mudry, C.}
\newblock \bibinfo{title}{Zero modes in the random hopping model}.
\newblock \emph{\bibinfo{journal}{Phys. Rev. B}} \textbf{\bibinfo{volume}{66}},
  \bibinfo{pages}{014204} (\bibinfo{year}{2002}).

\end{thebibliography}
\end{document}